%% file: main.tex
\documentclass[copyright,creativecommons]{eptcs}
\providecommand{\event}{GCM 2024} 

\usepackage{iftex}

\ifpdf
  \usepackage{underscore}         
  \usepackage[T1]{fontenc}        
\else
  \usepackage{breakurl}           
\fi

\usepackage{graphicx,color}
\usepackage{amsmath}
\usepackage{amssymb}
\usepackage{caption}
\usepackage{subcaption}
\usepackage[frozencache,cachedir=minted-cache]{minted}

\newminted{ocaml}{frame=single,framesep=5pt}
\newminted{text}{frame=single,framesep=5pt}

\title{An Encoding of Interaction Nets in OCaml}
\author{Nikolaus Huber
\institute{Uppsala University}
\email{nikolaus.huber@it.uu.se}
\and
Wang Yi 
\institute{Uppsala University}
\email{wang.yi@it.uu.se}
}
\def\titlerunning{An Encoding of Interaction Nets in OCaml}
\def\authorrunning{Huber and Yi}

\begin{document}
\maketitle

\begin{abstract}
Interaction nets constitute a visual programming language grounded in graph transformation. Owing to their distinctive properties, they inherently facilitate parallelism in the rewriting step. This paper showcases a simple and concise approach to encoding interaction nets within the programming language OCaml, emphasising correctness guarantees. To achieve this objective, we encode not only the interaction net primitives, but also Lafont's original type system.
\end{abstract}

\input{parts/introduction}
\input{parts/background}
\input{parts/implementation}
\input{parts/benchmarks}
\input{parts/conclusion}

\bibliographystyle{eptcs}
\bibliography{references}

\appendix 
\input{parts/app}

\end{document}

%% file: parts/introduction.tex
\section{Introduction}
\label{sec:introduction}

Interaction nets, introduced by Lafont in~\cite{lafont-inets}, are a model of computation based on graph rewriting. They have, among other things, been used as a basis for optimal~\cite{geometry-optimal-lambda,lamping-optimal} and efficient implementations~\cite{yale} of the $\lambda$-calculus. Interaction nets offer a number of desirable properties, such as locality of reduction, strong confluence, and Turing completeness. Owing to the locality of reduction, highly parallel implementations can be devised, while the confluence guarantees determinism. They have been studied both as a model for the dynamics of computation and also as a programming language itself. However, only a few implementations are still actively maintained. 

In this paper, we study a method of encoding interaction nets into the general purpose programming language OCaml. Unlike previous efforts, our focus is not on the efficiency of reduction, but rather on embedding  Lafont's original type system to provide stronger correctness guarantees. By utilising advanced type system features of the host language, we can statically enforce Lafont's typing discipline without relying on dynamic checks. 

To the best of our knowledge, this is the first encoding of interaction nets in OCaml, and the first attempt at embedding Lafont's type system into the type system of another language. Given OCaml's recent addition of native support for expressing parallelism, we therefore also explore whether interaction nets can serve as a general implementation scheme for parallel algorithms in the language.

The structure of this paper is as follows: In Section \ref{sec:background} we give an overview of the model of interaction nets together with the graphical notation we will be using throughout the paper, followed by a brief introduction to OCaml and an overview of previous work. In Section \ref{sec:implementation} we illustrate our method of encoding interaction nets into OCaml by way of showing a concrete example. In Section \ref{sec:benchmarks} we look at the runtime behaviour of different encoded nets, and investigate if they can make use of OCaml's support for parallelism. Finally, we give directions for future work and conclude in Section \ref{sec:conclusion}.

%% file: parts/background.tex
\section{Preliminaries}
\label{sec:background}

\subsection{Interaction nets}

\begin{figure}
    \begin{subfigure}[b]{0.3\textwidth}
        \centering
        \includegraphics[width=\textwidth]{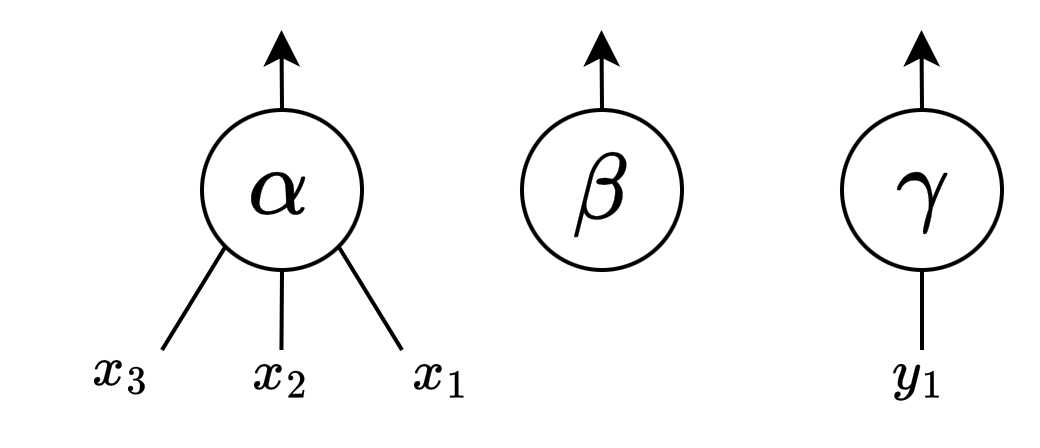}
        \caption{Example agents}
        \label{fig:agent-examples}
    \end{subfigure}
    \hfill
    \begin{subfigure}[b]{0.6\textwidth}
        \centering
        \includegraphics[width=\textwidth]{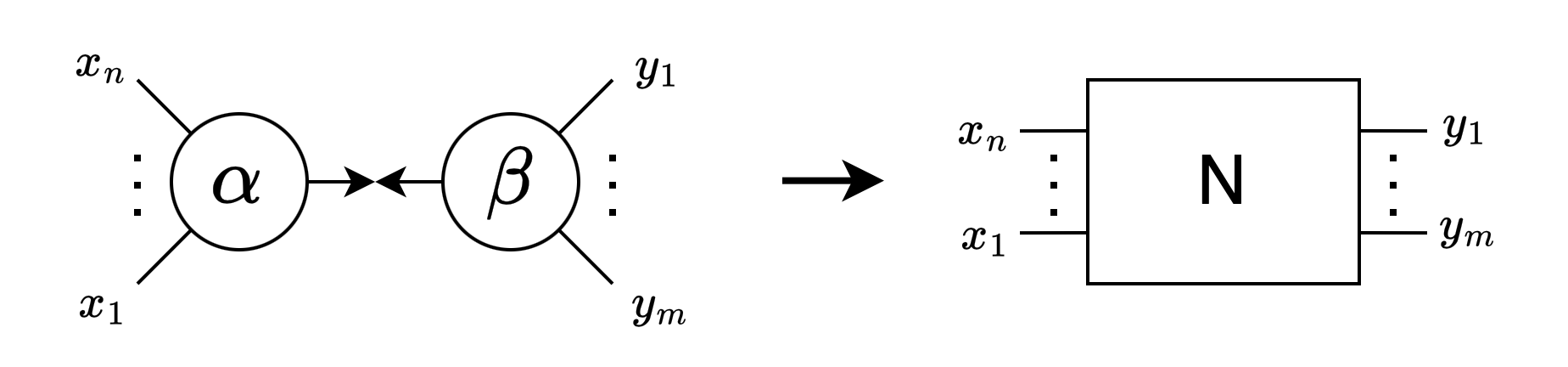}
        \caption{Example rule}
        \label{fig:rule-example}
    \end{subfigure}
    \caption{Graphical notations for agents and rules}
\end{figure}

We briefly recall the basic notions of interaction nets. For a more complete presentation, the interested reader is referred to~\cite{lafont-inets}. Interaction nets consist of the following elements: 

\begin{itemize}
    \item A set $\Sigma$ of \emph{symbols}. These symbols are used as labels for nodes, which are referred to as \emph{agents}. Each agent has a fixed set of \emph{ports}, through which they connect with other agents. An agent has exactly one distinguished \emph{principle port}, and a (possibly empty) set of \emph{auxiliary ports}. The number of auxiliary ports is fixed for each symbol. We assume the existence of a function $ar: \Sigma \rightarrow \mathbb{N}$, such that, for all $\alpha \in \Sigma$, $ar(\alpha)$ is the number of auxiliary ports associated with the symbol $\alpha$. Figure \ref{fig:agent-examples} illustrates the usual graphical notation for agents of three different symbols $\alpha$, $\beta$, and $\gamma$. The principle port is depicted as an arrow, auxiliary ports are numbered clockwise from the principle port. In the given example $ar(\alpha) = 3$, $ar(\beta) = 0$, and $ar(\gamma) = 1$. 
    \item A net $N$ is an undirected graph built from agents over the set $\Sigma$. Each edge of the graph connects two agent ports together, such that there is only one edge per port. A port that is not connected is called a \emph{free port}. The set of free ports of a net is called its \emph{interface}. 
    \item A set $\mathcal{R}$ of \emph{rules}. A pair of agents $(\alpha, \beta) \in \Sigma \times \Sigma$ is called an \emph{active pair} if they are connected through their principle ports (this is related to the concept of a \emph{redex} in term rewriting). The application of an interaction rule $((\alpha, \beta) \rightarrow N) \in \mathcal{R}$ replaces an active pair by a net $N$ in such a way that the interface of $N$ coincides exactly with the auxiliary ports of $\alpha$ and $\beta$. Intuitively, this means, that all auxiliary connections of the active pair must be connected to the new net, and no new free ports can be introduced in the process. Figure \ref{fig:rule-example} shows an example of the graphical notation for rules. 
\end{itemize}

Rewriting happens through the application of interaction rules to active pairs. Since every agent only has one distinct principle port, the rewriting process is local, and two different active pairs can be rewritten in parallel. Some additional requirements are necessary to guarantee determinism: For each pair $(\alpha, \beta)$ there can only be at most one rule in $\mathcal{R}$, and if there is no rule for a specific active pair it cannot be further reduced. Since there is no notion of orientation, rules are symmetric, meaning that $(\alpha, \beta) \rightarrow N$ and $(\beta, \alpha) \rightarrow N$ describe the same rule. With these restrictions in place it can be shown~\cite{lafont-inets}, that the reduction sequence is strongly confluent. 

\subsection{Typed interaction nets}

Already in the original paper~\cite{lafont-inets}, Lafont remarks that not all possible connections between agents have a valid semantic interpretation. In the next section, we will introduce agents representing boolean and integer values, as well as agents representing functions over those values. A function agent expecting a particular type at its principle port cannot be reduced when paired with an agent of the wrong type. Therefore, Lafont introduced a rudimentary type system, where each port of an agent is assigned both a \emph{value type}, and a \emph{polarity}. Value types can include integers, booleans, floats, lists, etc. 

\begin{figure}[th]
    \centering
    \includegraphics[width=0.3\linewidth]{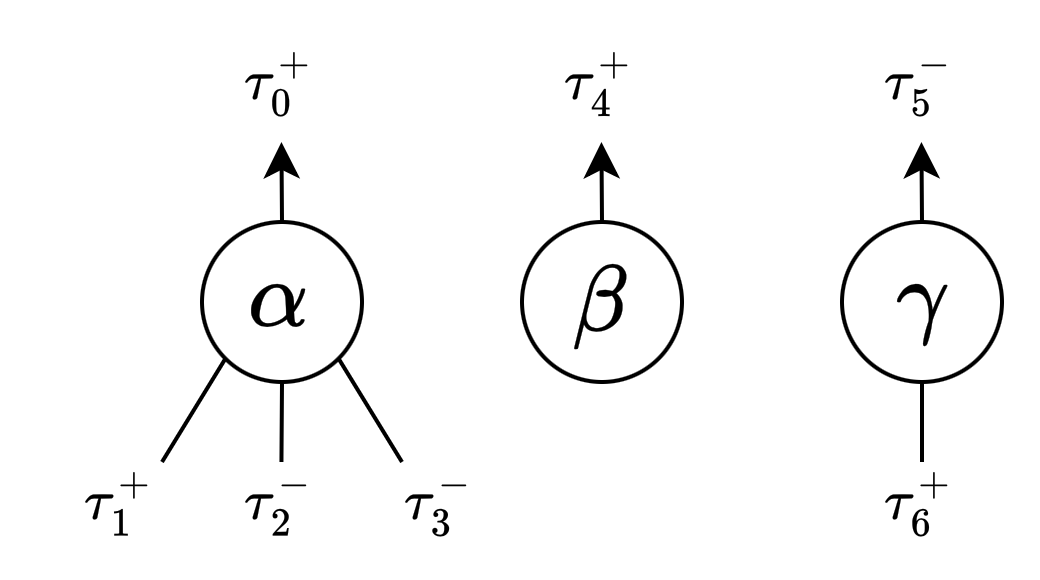}
    \caption{Examples of typed agents}
    \label{fig:typed-agents}
\end{figure}

The polarity encodes if a port is meant to be an input or an output. An example can be seen in Figure~\ref{fig:typed-agents}. It is up to convention if inputs receive positive, or negative polarity, as long as the assignment is consistent throughout. In this paper, we will follow the same convention as in~\cite{lafont-inets}, where inputs have negative, and outputs positive polarity. 

Only ports of the same type, but opposite polarity, can be connected to each other. Rules are well typed if their left-hand side is well typed (i.e., the principle ports of the active pair have the same type and opposite polarity), and if the net on the right-hand side is well typed, taking into account the interface types provided by the auxiliary connections of the active pair. 

\subsection{OCaml}

OCaml is an industrial-strength (primarily functional) programming language which arose as an extension of the Caml dialect of the ML language family. While it is a general-purpose language, it has a strong background in theorem proving (e.g., Coq~\cite{coq}), static analysis (e.g., Frama-C~\cite{frama-c}), and formal methods software. Interestingly, even though it is a language used in industry, it only recently added native support for expressing parallel evaluation~\cite{parallel-ocaml}. 

The basic unit of parallelism in OCaml is called \emph{domains}. A single domain is created at the start of a program, and more domains can be spawned later on. Domains map directly to operating system threads, and are therefore costly to create and tear down. It is thus recommended to not spawn more domains than cores available on the local processor. While the standard library of OCaml provides the \texttt{Domain} module, it only offers low-level primitives for managing domains. This opens the possibility of providing higher-level libraries to be developed outside the core compiler. Different such libraries have been developed, all of them with different design trade-offs and different APIs. In order not to be too restricted by the choice of library, we will show our encoding method against a simplified API, which can be provided by different libraries: 

\begin{ocamlcode}
type pool 
type 'a promise
type 'a resolver
val resolve : 'a resolver -> 'a -> unit 
val await : 'a promise -> 'a 
val block : 'a promise -> 'a 
val make_future : unit -> 'a promise * 'a resolver 
val create_pool : int -> pool 
val run_async : pool -> (unit -> unit) -> unit 
\end{ocamlcode}

Type \texttt{pool} describes the pool of worker threads, it is created by a call to \texttt{create_pool} with the number of desired workers as an argument. Types \texttt{'a promise} and \texttt{'a resolver} are two ends of a simple communication primitive. They are always created together by a call to \texttt{make_future}. A value of type \texttt{'a promise} acts as a placeholder for a value of type \texttt{'a}, which will later be supplied through the corresponding \texttt{resolver} by calling the \texttt{resolve} function. Function \texttt{await} takes a promise and returns its value once it is resolved. Function \texttt{block} does the same. However, it blocks the currently running thread until the promise is resolved. Finally, the function \texttt{run_async} takes a pool and a continuation, and puts that continuation into the work-queue of the pool for asynchronous execution. Such an API can be provided by different already existing libraries, the one we chose for this paper is called Moonpool~\cite{moonpool}. For those unfamiliar with the OCaml language, we give a short overview of the syntax in appendix \ref{app:ocaml}.

\subsection{Previous work and contributions} 

A number of different evaluators/compilers for interaction nets have been developed. Some have built upon each other, such as AMINE~\cite{pinto-amine}, PIN~\cite{hassan-pin}, INET~\cite{hassan-compilation}, and amineLight~\cite{mackie-lightweight}. A comparison of these is given in~\cite{sato-phd}, and the most recent and still actively maintained version of this lineage is available in the inpla project~\cite{inpla}. 

Due to the inherent possibility of parallel reduction, running interaction nets on GPUs was investigated in~\cite{Jiresch-2014}. The code for the project is still available~\cite{ingpu}. However, it has not been updated in 13 years and needs all rules encoded as custom CUDA kernels. 

Encoding interaction nets in a functional programming language was done in~\cite{Kahl-2015}, where interaction nets were embedded into Concurrent Haskell~\cite{concurrent-haskell}. The code for this evaluator is available~\cite{hinet}, but it also has not been updated since 2015, and requires a rather old version of  Haskell to still compile. 

Our method is closest to the embedding into Haskell. However, we have taken multiple different design choices. In the Haskell encoding, the arity constraint for each label is not enforced by the type system, as ports are encoded as a list of directed references to other agents. In our encoding, not all connections are implemented as references, and utilising a dedicated constructor for each agent the type system guarantees the arity constraints. The Haskell encoding uses polarity to figure out the direction for each reference, i.e., which side will provide the reference, and which side waits until the reference is provided. It relies on a heuristic to figure out the polarity of each port. In this paper, we bypass the inference of polarities, and instead assume that they are given as part of the agent definitions. In the Haskell encoding, polarities are carried around as concrete values, and are dynamically checked. We encode polarities in the type system of OCaml directly, therefore preventing the construction of incorrect nets already during compile time, and removing the need for dynamic checks. 

%% file: parts/implementation.tex
\section{Encoding}
\label{sec:implementation}

In this section, we showcase the encoding of the different constituent parts of interaction nets into OCaml. We start by defining a dedicated type for agents, then show how we encode rule application. After introducing agent attributes, we then show how we embedded Lafont's original type system into the type system of OCaml.  

\subsection{Encoding agents} 

\begin{figure}[t]
    \centering
    \includegraphics[width=0.4\linewidth]{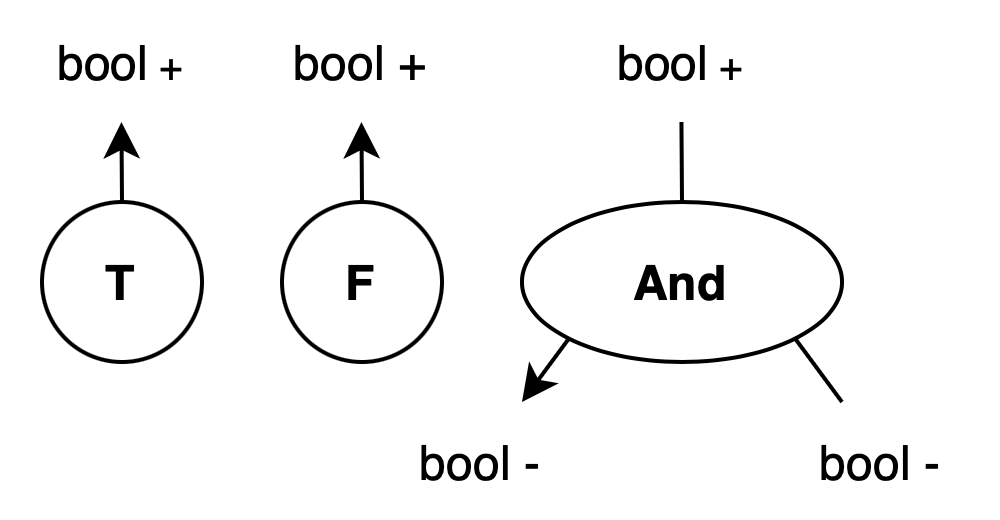}
    \caption{Boolean agents}
    \label{fig:bool-agents}
\end{figure}

There are many ways of how agents might be encoded within a given language. In the interaction net to C compiler presented in~\cite{hassan-compilation}, agents are encoded as a structure, where the first field relates to the label of the respective agent, while the second encodes the ports as an array of references to other agents. The encoding of interaction nets in Haskell presented in~\cite{Kahl-2015} uses a similar approach, where each agent is described as a record with a field for the label and a field for the list of connections to other agents. These connections make use of mutable variables, as introduced by Concurrent Haskell~\cite{concurrent-haskell}. Indeed, the way that mutual connections of agents are encoded seems to be one of the most important differences between different interaction net evaluators. In his PhD thesis~\cite{sato-phd}, Sato gives a nice overview of these different encodings. According to his nomenclature, our encoding method follows the principle of \emph{single link encoding}, which seems to also be the basis of the Haskell encoding. 

For a given interaction net system with a label set $\Sigma$, we create a \emph{variant data type} with one constructor case per symbol. As a simple example, we will start by encoding the agents shown in Figure~\ref{fig:bool-agents}: 

\begin{ocamlcode}
type agent = 
    | T
    | F 
    | And of agent * agent 
\end{ocamlcode}

Each symbol $\alpha \in \Sigma$ is translated to one constructor case in the type \texttt{agent}, using $\alpha$ as the label, and carrying $ar(\alpha)$ agents as attributes (encoding the appropriate number of auxiliary ports). When creating a value of type \texttt{agent} by using one of these constructors, the resulting value represents the principle port of the respective agent. Every auxiliary port used as an attribute is a connection to the principal port of another agent. 

\begin{figure}[th]
    \begin{subfigure}[b]{0.4\linewidth} 
        \centering 
        \includegraphics[width=0.8\linewidth]{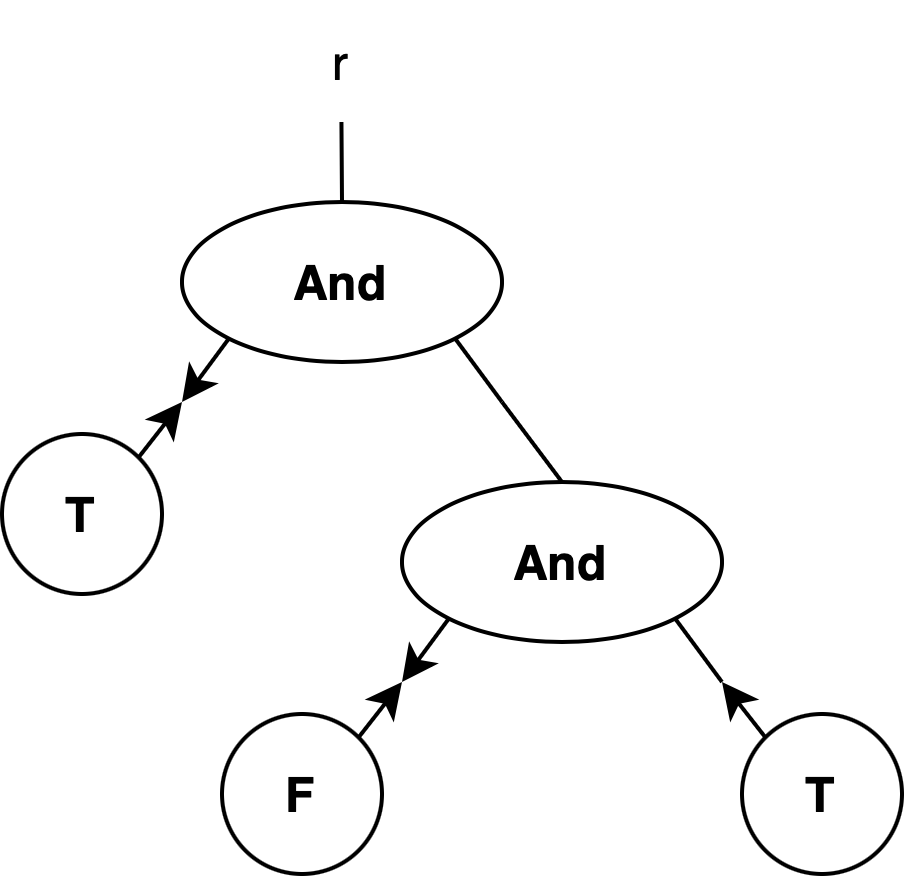} 
        \caption{Using common connection}
        \label{fig:aux-conn}
    \end{subfigure}
    \begin{subfigure}[b]{0.6\linewidth} 
        \centering 
        \includegraphics[width=0.8\linewidth]{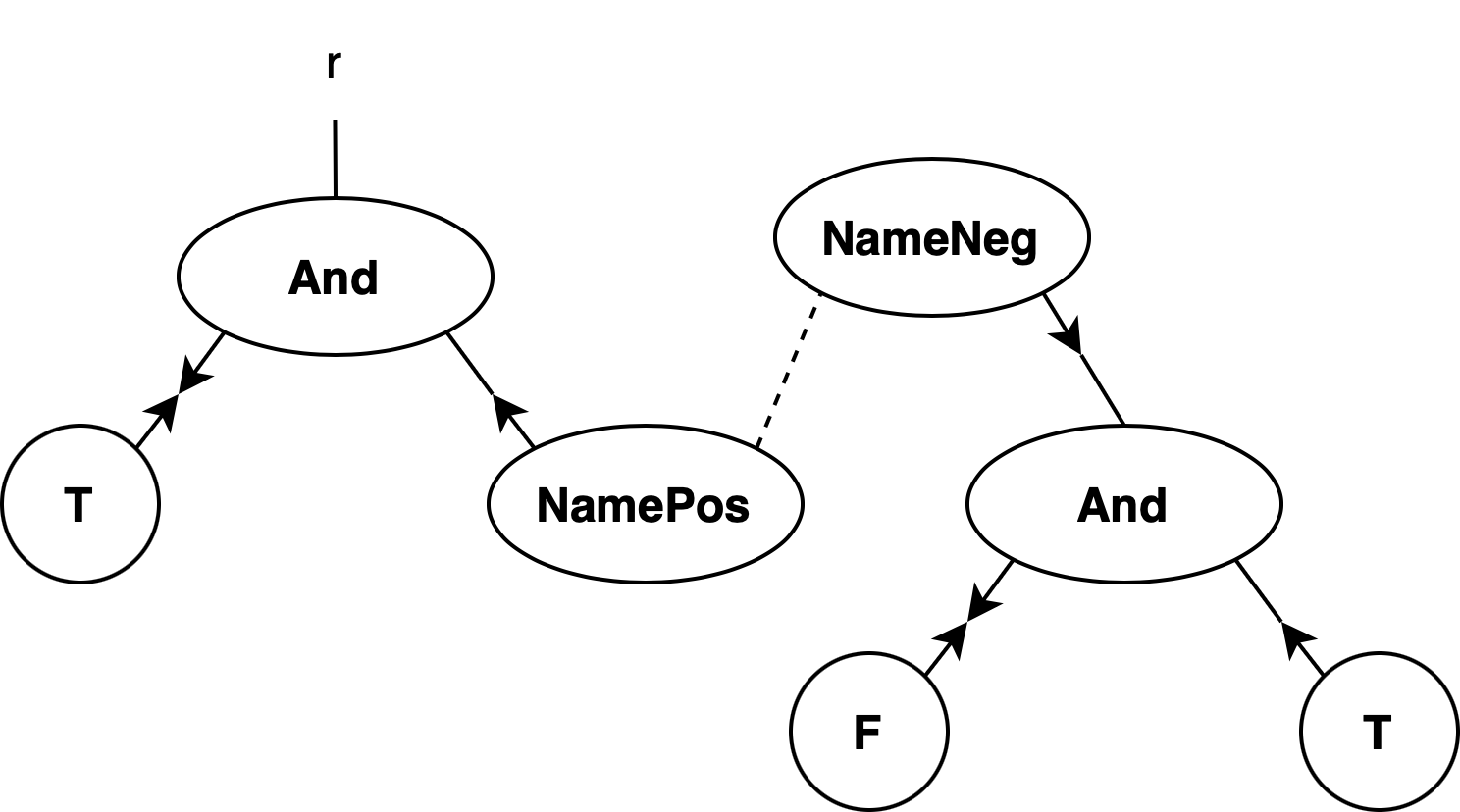}
        \caption{Using name nodes}
        \label{fig:conn-name-nodes}
    \end{subfigure}
    \caption{Connection between auxiliary ports}
\end{figure}

Nets are created by simply applying constructors recursively to each other. However, encoding a net in this way does not allow for two auxiliary ports to be connected directly, so the net in Figure~\ref{fig:aux-conn} could not be encoded in this way. To allow the mutual connection of auxiliary ports, we break the link and create two agents with new symbols \texttt{NamePos} and \texttt{NameNeg} (see Figure \ref{fig:conn-name-nodes}). We utilise the information about the polarity of the connection in order to decide which port needs to be connected to which name agent. We can therefore extend our \texttt{agent} type accordingly: 

\begin{ocamlcode}
type agent = 
    | T
    | F 
    | And of agent * agent 
    | NamePos of agent promise 
    | NameNeg of agent resolver 
\end{ocamlcode}

We use the same mechanism for creating connections for the interface of the initial network (i.e., the network to be rewritten). We also introduce a function to create these pairs of name agents: 

\begin{ocamlcode}
let new_name () = 
    let promise, resolver = make_future () in 
    NamePos promise, NameNeg resolver 
\end{ocamlcode}

\subsection{Encoding rule application}

\begin{figure}[ht]
    \begin{subfigure}[b]{0.4\linewidth} 
        \centering 
        \includegraphics[width=0.8\linewidth]{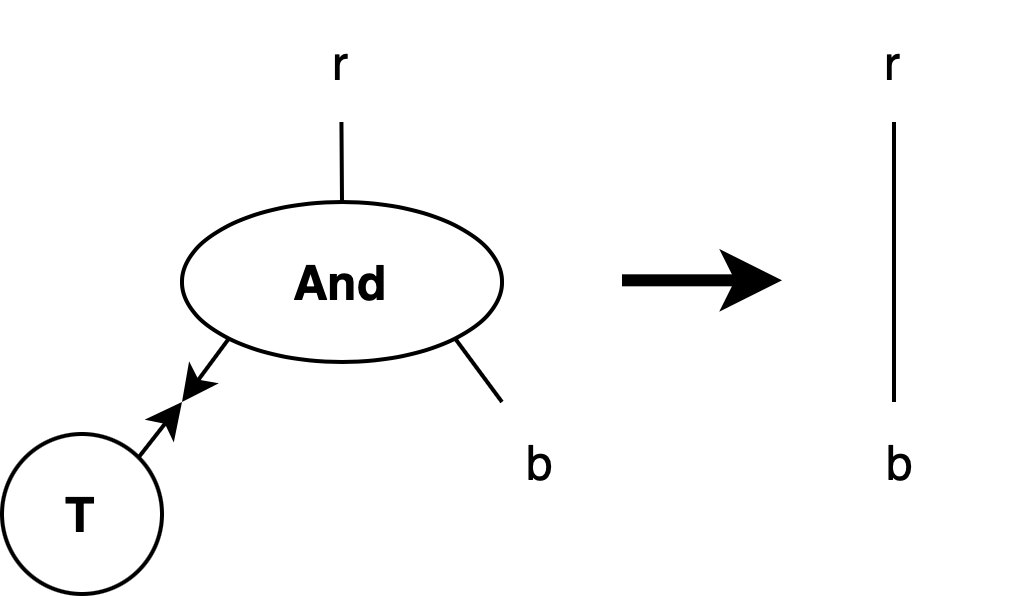} 
        \caption{Rule T $\rightarrow\leftarrow$ And}
        \label{fig:rule-true-and}
    \end{subfigure}
    \hfill
    \begin{subfigure}[b]{0.6\linewidth} 
        \centering 
        \includegraphics[width=0.8\linewidth]{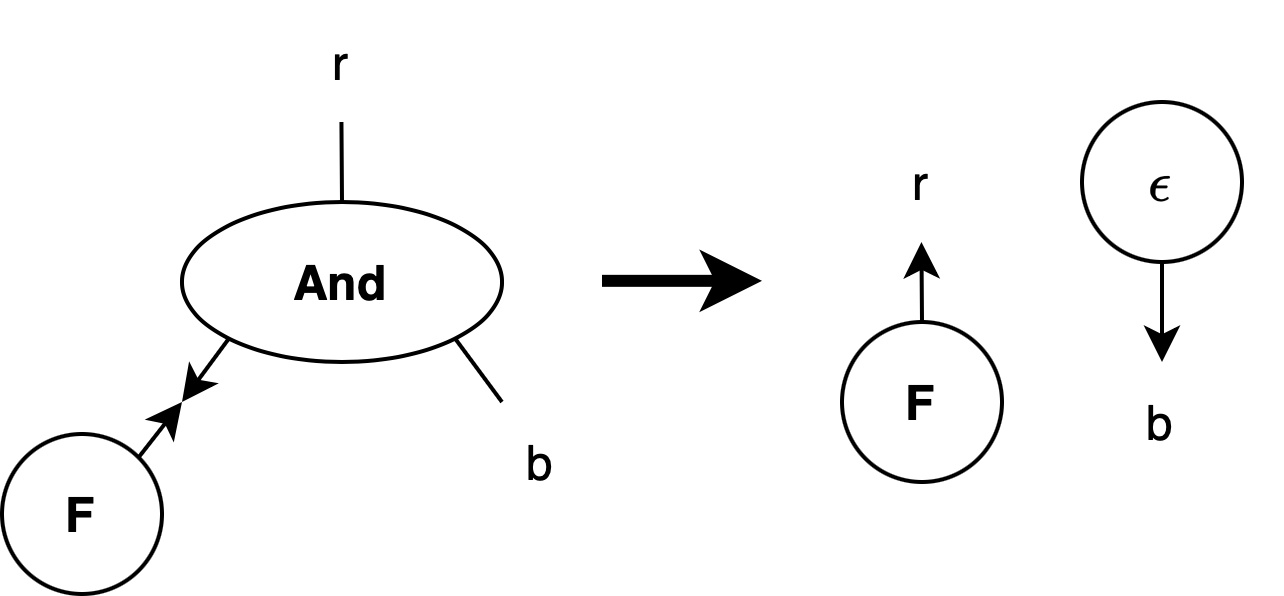}
        \caption{Rule F $\rightarrow\leftarrow$ And}
        \label{fig:rule-false-and}
    \end{subfigure}
    \caption{Rules for conjunction}
    \label{fig:conj-rules}
\end{figure}

Apart from agents, we also need to encode rules. All rules are collected and translated together into a function \texttt{apply_rule}, which internally uses pattern matching to match the different active pairs. So for the two rules shown in Figure \ref{fig:conj-rules} we get: 

\begin{ocamlcode}
let rec apply_rule a1 a2 = match a1, a2 with 
    (* using an or-pattern to list both orientations *)
    | T, And (r, b)
    | And (r, b), T         -> b -><- r 
    | F, And (r, b)                     
    | And (r, b), F         -> ignore b; F -><- r 
    | NamePos v, a  
    | a, NamePos v          -> await v -><- a
    | NameNeg v, a 
    | a, NameNeg v          -> resolve v a
    (* match cases must be exhaustive in OCaml *)
    | _, _                  -> failwith "No rule for this pair"

(* custom infix operator to run rewriting asynchronously in thread pool *)
and ( -><- ) a1 a2 = run_async pool (fun _ -> apply_rule a1 a2) 
\end{ocamlcode}

Since all rules are symmetric in the constituents of the active pair, we have to list both orientations. OCaml has a shorthand syntax for when two cases have the same right-hand side, as long as we name the attributes of matched constructors the same. The functions \texttt{apply_rule} and \texttt{-><-} are mutually recursive, the latter uses OCaml's support for custom infix operators and just puts the application of \texttt{apply_rule} into the pool of threads to run asynchronously with all other rewritings. While this is not strictly necessary, it simplifies the code in the \texttt{apply_rule} method considerably. We assume that the variable \texttt{pool} has been globally defined at the start of the program. To satisfy the exhaustiveness of the pattern match, we need to include a catch-all case (\texttt{_, _}), which just fails with an error message. As we will see in Section \ref{sec:polarity}, once we introduce Lafont's typing discipline, this catch-all case will become unnecessary. 

The careful reader will have realised that we have not introduced the agent with label $\epsilon$ as shown in Figure~\ref{fig:rule-false-and}. This is a special agent, usually referred to as the \emph{delete agent}, which just recursively deletes the net connected to it. As OCaml is a managed language, we can leave this deletion process to the garbage collector. However, the compiler will complain if we name a port on the left-hand side of a case and not use it on the right-hand side. We therefore use the OCaml function \texttt{ignore} to mark that we do not care about a particular value. Another option would have been to use the special \texttt{_} pattern in order not to give a name to the unused port: 

\begin{ocamlcode}
... 
    | F, And (r, _) 
    | And (r, _), F    -> F -><- r
... 
\end{ocamlcode}

\subsection{Agent attributes}

\begin{figure}[ht]
    \centering
    \includegraphics[width=0.6\linewidth]{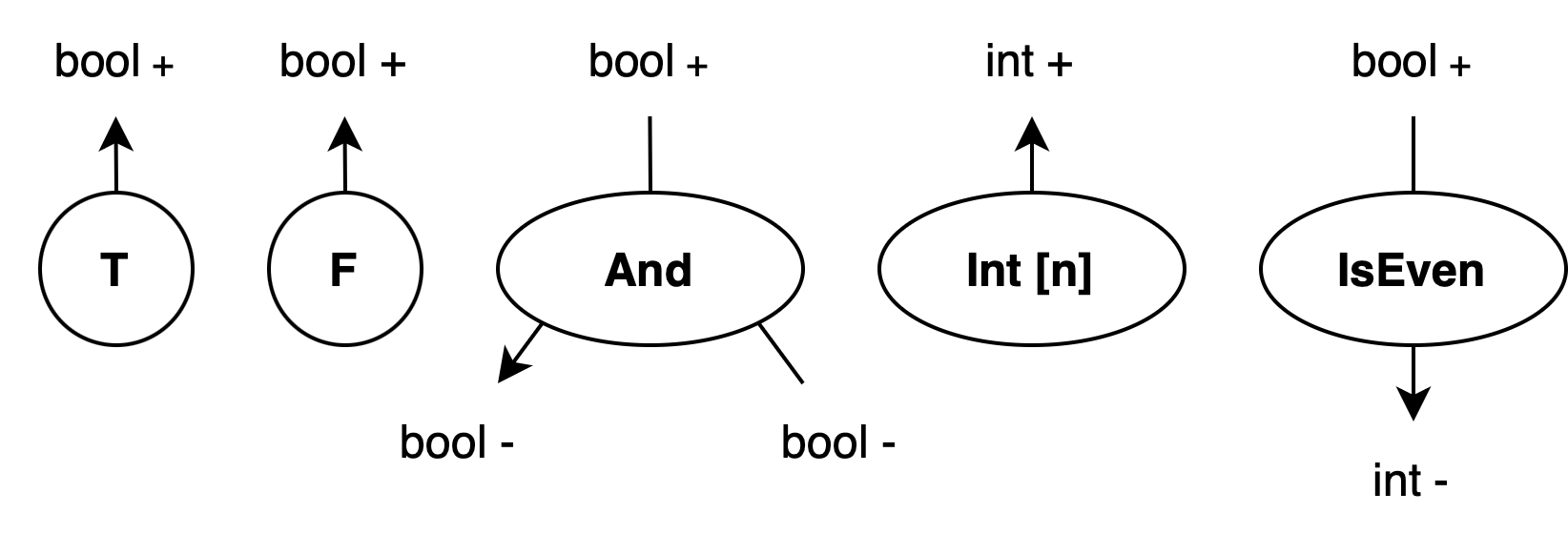}
    \caption{Boolean agents extended with integer agents}
    \label{fig:bool-int-agents}
\end{figure}

Representing booleans with two different labels works fine. However, with numbers such a unary encoding becomes already quite tedious and inefficient. Therefore, we extend our definition of agents, allowing them to carry attributes, i.e., values. As shown in~\cite{hassan-compilation}, this does not contradict the strong confluence of the reduction. Figure \ref{fig:bool-int-agents} shows an extension of our previous label set with \texttt{Int} agents, carrying a value of their namesake, as well as an agent \texttt{IsEven}. Encoding agents with attributes in OCaml is straightforward, as we can just extend the attributes of the respective constructor: 

\begin{ocamlcode}
type agent = 
    | Int of int 
    | IsEven of agent 
    | T
    | F 
    | And of agent * agent 
    | NamePos of agent promise  
    | NameNeg of agent resolver 
\end{ocamlcode}

\begin{figure}[ht]
    \centering
    \includegraphics[width=0.5\linewidth]{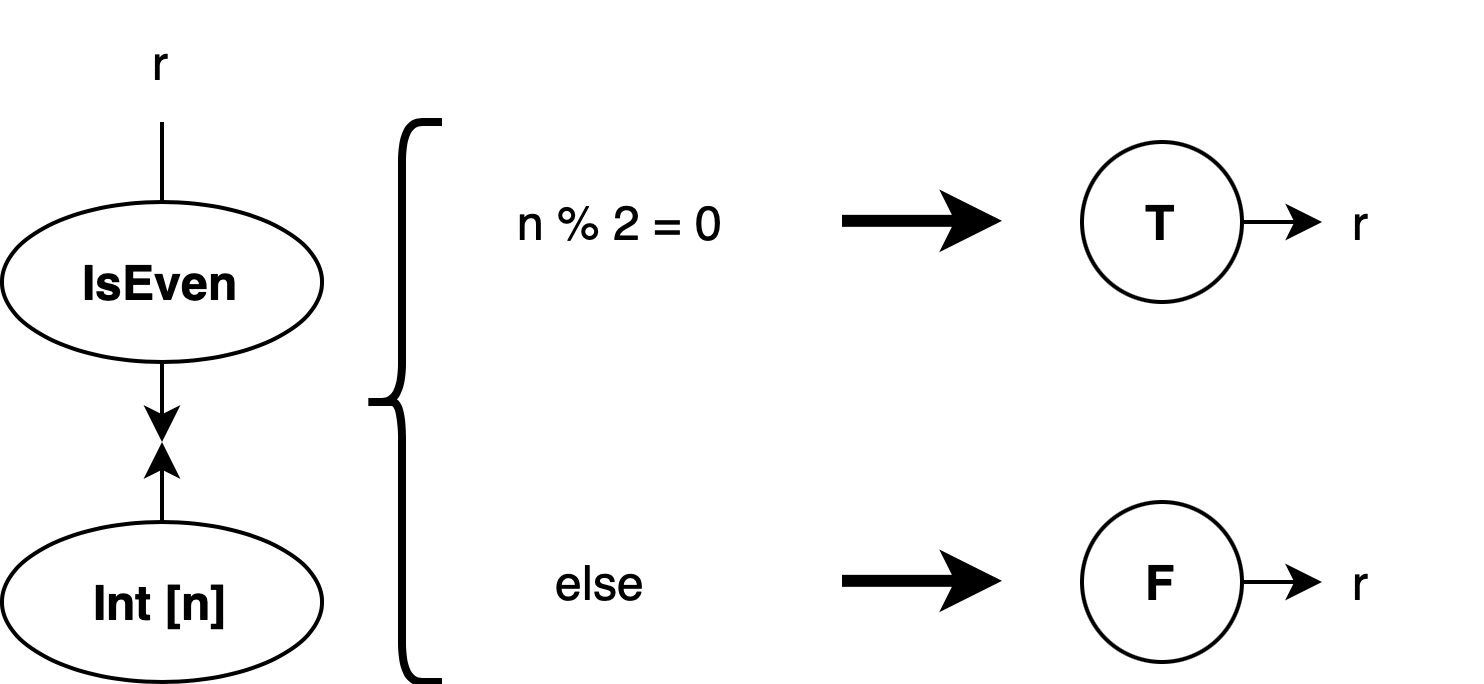}
    \caption{Rule Int[n] $\rightarrow\leftarrow$ IsEven}
    \label{fig:rule-int-iseven}
\end{figure}

We can use the attributes of the agents on the left-hand side of a rule to define guarded right-hand sides, as shown in Figure \ref{fig:rule-int-iseven}. Encoding such rules in \texttt{apply_rule} is easy, as OCaml allows match cases to be guarded by a boolean expression: 

\begin{ocamlcode}
let rec apply_rule a1 a2 = match a1, a2 with 
    ... 
    | IsEven r, Int n 
    | Int n, IsEven r when n mod 2 = 0    -> T -><- r 
    | IsEven r, Int _ 
    | Int _, IsEven r                     -> F -><- r
    ... 
\end{ocamlcode}

\subsection{Type compatibility} 

With the introduction of integers into our example, we now also have to consider the types of ports. Indeed, the way we have defined agents through an ordinary variant type does not enforce this, so the OCaml compiler will happily accept \texttt{And (r, Int 0)} as a valid expression, even though \mbox{\texttt{Int 0}} is not a boolean. We could, of course, try and include a deep embedding of the type of an agent by doing something similar to the following:

\begin{ocamlcode}
type ty = 
    | Int 
    | Bool 

type agent = 
    | Int of ty * int 
    | IsEven of ty * agent 
    | T of ty 
    | F of ty 
    | And of ty * agent * agent 
    | ... 
\end{ocamlcode}

However, encoding the type in this way incurs an overhead both in execution time and code conciseness, as we would need to include runtime checks at appropriate positions to check for type compatibility. 

It would be nicer, if we could track the type of an agent in OCaml's type system directly, in a way that would prevent the construction of such faulty nets in the first place. Luckily, OCaml offers exactly this in the form of \emph{generalised algebraic data types} (GADTs). A GADT is similar to a parameterised variant type, but it allows expressing constraints on the type variable for each constructor case individually. For our example, the agent type can be defined in the following way: 

\begin{ocamlcode}
type _ agent =
  | Int : int -> int agent 
  | IsEven : bool agent -> int agent  
  | T : bool agent 
  | F : bool agent 
  | And : bool agent * bool agent -> bool agent  
  | NamePos : 'a agent promise -> 'a agent 
  | NameNeg : 'a agent resolver -> 'a agent
\end{ocamlcode}

The \texttt{:} after the constructor label indicates the definition of a GADT. The \texttt{\textunderscore} before \texttt{agent} is an \emph{anonymous type variable}, as it does not show up in the definition of the type \texttt{agent} itself. However, for each constructor case this type variable can be constrained to a concrete type, thereby limiting the allowed agents at specific auxiliary ports. For example, the constructor \texttt{And} expresses, that both auxiliary agent connections must be of type \texttt{bool}, the principle port has type \texttt{bool} as well (which, by lucky syntactic coincidence, follows after the \texttt{->}). With this in place, the OCaml compiler will now reject \texttt{And (r, Int 0)} as a valid net.

The type checker usually needs additional information when GADTs are paired with recursive functions (such as \texttt{apply_rule}). In general, if a recursive function uses pattern matching, the type checker fails to unify the type variable with different types in the different cases. It also does not allow for a recursive call to use different type variables than the outer call. This can be circumvented by providing explicit type annotations that mark the function as polymorphic in the type parameter of the types of its input: 

\newpage

\begin{ocamlcode}
let rec apply_rule : type a. a agent -> a agent -> unit = 
    fun a1 a2 -> ... 

and ( -><- ) : type a. a agent -> a agent -> unit  = 
    fun a1 a2 -> ...
\end{ocamlcode}

Intuitively, the annotation \texttt{type a} can be read as a for-all quantifier.

\subsection{Polarity}
\label{sec:polarity}

Type compatibility between ports is only one part of Lafont's type system. It does not prevent us from creating active pairs that do not fit together in terms of their polarity, so currently, the compiler will accept \texttt{Int 0 -><- Int 1} as valid. To encode the polarity, we can introduce a second anonymous type variable for the definition of the polarity type. If we, for now, assume that there are defined types \texttt{pos} and \texttt{neg}, we can express the agent type in the following way:

\begin{ocamlcode}
type (_, _) agent =
  | Int : int -> (int, pos) agent 
  | IsEven : (bool, neg) agent -> (int, neg) agent  
  | T : (bool, pos) agent 
  | F : (bool, pos) agent 
  | And : (bool, neg) agent * (bool, pos) agent -> (bool, neg) agent  
  | NamePos : ('a, pos) agent promise -> ('a, pos) agent 
  | NameNeg : ('a, pos) agent resolver -> ('a, neg) agent
\end{ocamlcode}

The astute reader will have surely realised, that all auxiliary ports in the above definition have opposite polarity to what they had in their graphical depiction in Figure \ref{fig:bool-int-agents}. This change in polarity is due to the fact that we need to think in terms of the polarity of the agents that will be connected to these ports, therefore we need to invert them. With these types in place, we can refine the definition of \texttt{apply_rule}: 

\begin{ocamlcode}
let rec apply_rule : type a. (a, pos) agent -> (a, neg) agent -> unit = 
  fun a1 a2 -> match a1, a2 with 
  | T, And (r, b)                     -> b -><- r 
  | F, And (r, b)                     -> ignore b; F -><- r 
  | Int n, IsEven r when n mod 2 = 0  -> T -><- r 
  | Int _, IsEven r                   -> F -><- r  
  | T, If (r, t, _)                   -> t -><- r 
  | F, If (r, _, e)                   -> e -><- r
  | NamePos v, a                      -> await v -><- a  
  | a, NameNeg v                      -> resolve v a

and ( -><- ) : type a. (a, pos) agent -> (a, neg) agent -> unit = 
  fun a1 a2 -> run_async pool (fun _ -> apply_rule a1 a2)
\end{ocamlcode}

Interactions can only happen between agents that have the same type but opposite polarity. The choice of having the first agent be positive, and the second negative is arbitrary, and we could have introduced them just as well the other way around. 

With these type annotations in place, the OCaml compiler will now also inform us, that the catch-all case is superfluous, as we have accounted for all possible active pair patterns. It also means that only one orientation of the agents of an active pair is permitted in the pattern match any more, so we can shorten the code quite a bit. 

In it worth noticing, that the OCaml compiler performs not only exhaustiveness checks (see \ref{app:ocaml}), but also checks for overlaps in the case patterns. The order of cases matters (e.g., in the above example it is important that the case with the \texttt{when} guard is listed first), so later case patterns can generalise prior ones, but if two cases are overlapping completely, the OCaml compiler would complain. This further increases confidence, that the encoding is correct. 

It remains to define the two types \texttt{pos} and \texttt{neg}. In principle, any two types that cannot be unified would work, so we could make \texttt{pos} a synonym for \texttt{int} and \texttt{neg} a synonym for \texttt{float}, for example. However, since we never create values of these types (they are only used as \emph{tags} in the type definition), the most idiomatic choice is an \emph{uninhabited} or \emph{empty type}. OCaml allows the definition of such a type as a variant without constructors: 

\begin{ocamlcode}
type pos = | 
type neg = | 
\end{ocamlcode}

Since they are defined as two separate types, the type checker can never unify them. We could have, of course, encoded the polarity dynamically as another attribute of all agent constructors again, and put checks at relevant places (e.g., at agent construction and \texttt{apply_rule} applications). This would have come at an additional cost in both execution time and code size again. By tracking the polarity in the type system we incur neither of these, since types are removed by the compiler after type checking, and the compiler will reject expressions such as \texttt{Int 0 -><- Int 1} already at compile time. 

%% file: parts/benchmarks.tex
\section{Benchmarks}
\label{sec:benchmarks}

\begin{figure}[p]
    \centering 
    \includegraphics[width=0.6\textwidth]{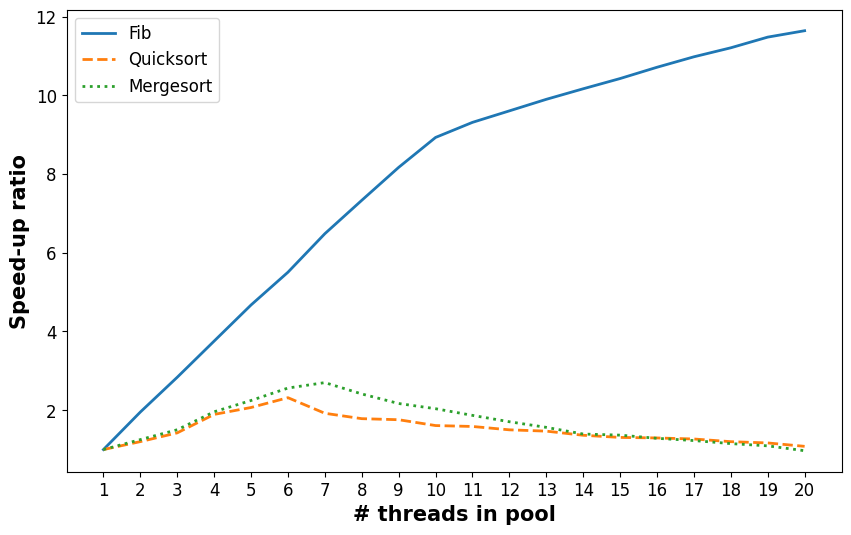}
    \caption{Relative speed-up factors for different pool sizes}
    \label{fig:speedup}
\end{figure}

\begin{figure}[p]
    \centering 
    \includegraphics[width=0.6\textwidth]{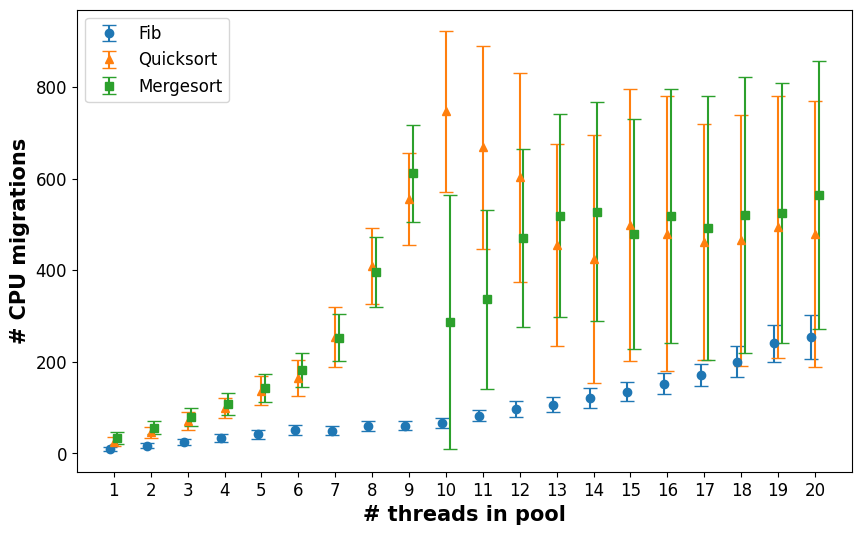}
    \caption{Number of CPU migrations for different pool sizes}
    \label{fig:cpu-migrations}
\end{figure}

\begin{figure}[p]
    \begin{subfigure}[b]{0.5\textwidth}
        \centering
        \includegraphics[width=0.8\textwidth]{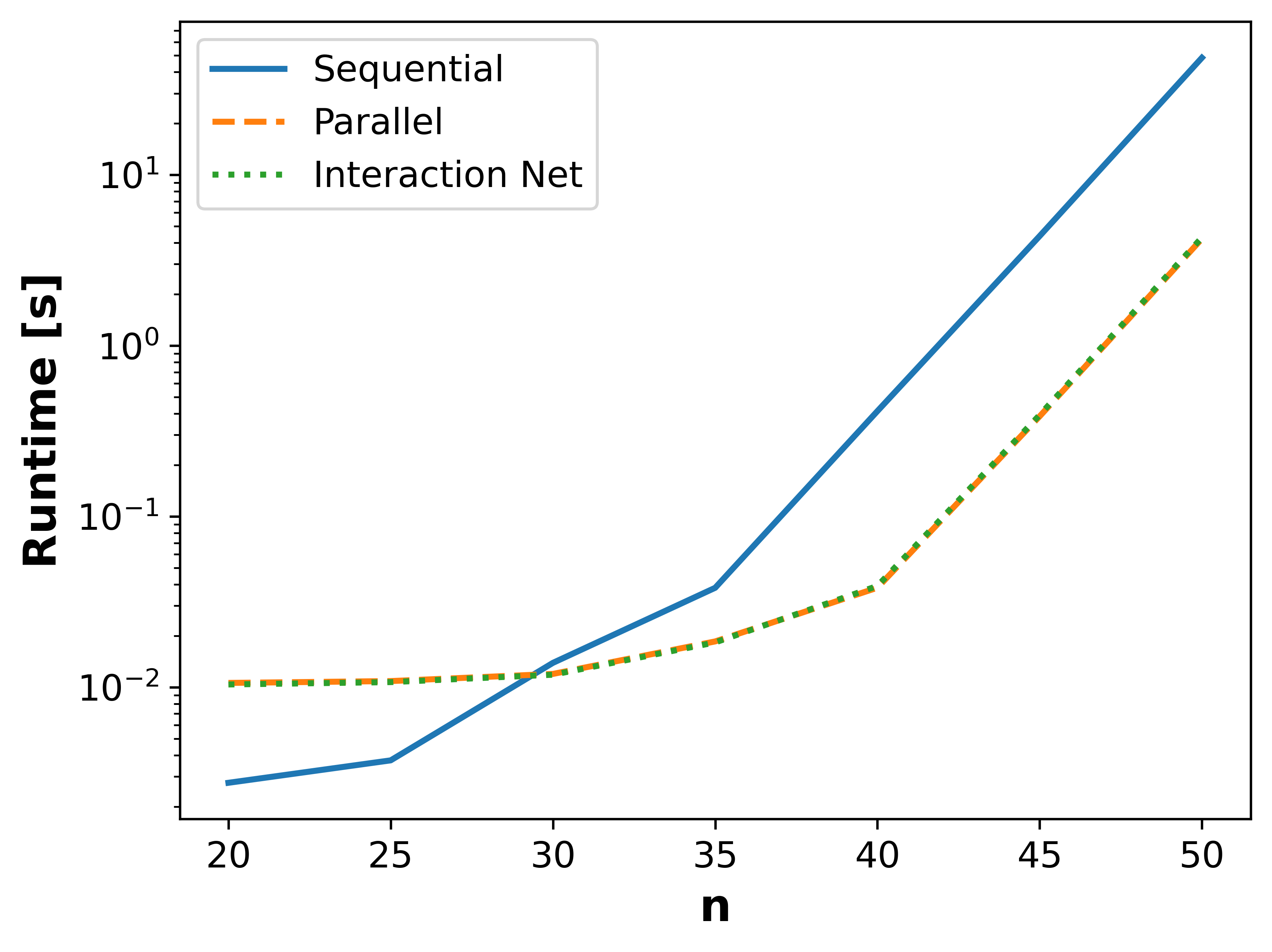}
        \caption{Runtime compared to handwritten OCaml}
        \label{fig:fib-results}
    \end{subfigure}
    \begin{subfigure}[b]{0.5\textwidth}
        \centering
        \includegraphics[width=0.8\textwidth]{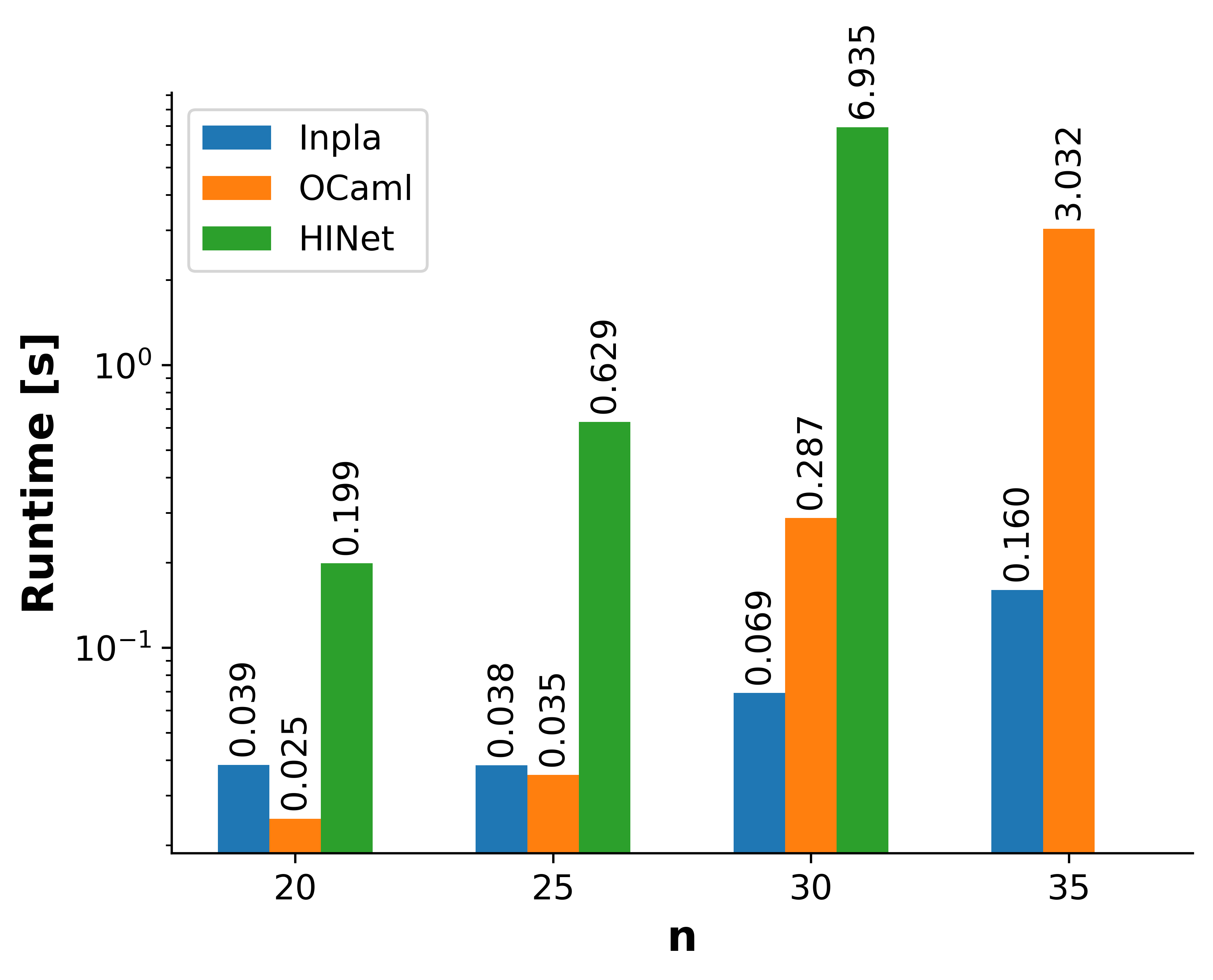}
        \caption{Tool comparison}
        \label{fig:tool-comparison}
    \end{subfigure}
    \label{fig:benchmarks-fib}
    \caption{Fibonacci benchmark}
\end{figure}

While the main goal of this paper is to express a clean, concise, and correct encoding of interaction nets into OCaml, it is of course interesting to look at the runtime behaviour of the encoded nets as well. To this end, we have encoded three example nets, taken from the inpla project~\cite{inpla}. Specifically, we have encoded nets for calculating Fibonacci numbers, and sorting lists of integers with Quicksort and Mergesort. Our code is available online~\cite{ocaml-inets}. 

All examples were run on a $3.70\;\text{GHz}$ Intel Core i9 processor with 10 cores and 2 hardware-threads per core. For each measurement we average over 100 runs. The first question is, of course, if the interaction net encoding can make use of the available cores. Figure \ref{fig:speedup} shows the relative speed-up ratios when run on pools of different sizes. It is important to notice that the library we are using (Moonpool~\cite{moonpool}) works in terms of thread pools, not domain pools. There is always exactly one domain pool created at the start of the program (with the recommended number of domains, i.e., the number of cores/hardware-threads available on the local processor), the library then creates pools of worker threads which share these domains. There are a couple of interesting observations: 

Fibonacci scales extremely well with regard to the size of the provided pool, while Quicksort's and Mergesort's performances only marginally increase, before they deteriorate for higher thread counts. There are several possible reasons for this. As a first indicator, we can look at the average amount of CPU migrations (threads moving from one CPU to another) for different thread counts, as shown in Figure \ref{fig:cpu-migrations}. The average number of CPU migrations for Fibonacci increases only slightly for higher thread counts, while showing exponential increases for both sorting algorithms until the number of physical cores is reached. We can also see that the average number of CPU migrations for both sorting algorithms fluctuates significantly for higher thread counts, as indicated by the standard deviations. While this gives a possible explanation for the lack of relative speed-up, further investigation is needed to look at other performance indicators such as cache misses and garbage collection behaviour. 

The speed-up ratios are relative to the performance of the sequential execution, they do not allow to reason about absolute performance. We have therefore also compared the performance of the Fibonacci encoding against the sequential and a handwritten parallel OCaml implementation. The results are shown in Figure \ref{fig:fib-results}. Parallelism incurs a price in terms of runtime, so for small inputs ($\leq 20$) we revert to the sequential algorithm for both the interaction net encoding and the parallel implementation. This shows another advantage of embedding interaction nets in a host language, as such tricks can easily be encoded.  

It is interesting to notice that the interaction net encoding is fairly on par with the handwritten parallel code. This suggests, that at least for some algorithms, interaction nets can be a valid candidate for implementing parallel algorithms with fine-grained parallelism in OCaml. 

For completeness, Figure \ref{fig:tool-comparison} compares the runtime of the Fibonacci net for different inputs $n$ in our encoding with two other interaction net evaluators. This comparison is not entirely fair, as they use vastly different strategies for running the provided net. Inpla~\cite{inpla} compiles interaction nets into bytecode for a low-level virtual machine dedicated to the execution of interaction nets. HINet~\cite{hinet} encodes interaction nets in Haskell dynamically, and then interprets these. It is therefore not surprising, that it is slower than the two other approaches. For the last test input, HINet did not produce an output within 10 minutes, and was therefore stopped. 

%% file: parts/conclusion.tex
\section{Conclusion}
\label{sec:conclusion}

In this paper, we presented a method of encoding interaction nets in the programming language OCaml. Our main focus was on an encoding of both the interaction net primitives (agents, nets, rules), as well as Lafont's type system. 

Compared to previous work, our encoding offers stronger guarantees of correctness. By defining a variant type with a constructor for each symbol, the type system guarantees the arity constraint and types of agent attributes. Using a GADT we can express both the value type and polarity of each port so that port compatibility is already checked at compile-time. Furthermore, it allows the OCaml type checker to prove the exhaustiveness of the rewriting step (i.e., that all possible active-pair patterns are accounted for). Since the encoding is purely done on the level of types, no additional runtime cost is incurred. 

Experiments indicate that the encoding of certain algorithms can make use of the inherent parallelism. Others seem to be constrained by the scheduling of individual threads on the available cores. More investigation is needed regarding different performance counters and optimal scheduling of the rewriting process.  

Another direction of future work relates to the choice of library to provide the primitives. As described in the introduction, the idea behind OCaml's low-level primitives in the standard library is that different (opinionated) higher-level libraries can be developed on top of it. The library we have used in this paper is only one possible option, and it would be interesting to see, if the choice of library has a significant impact on the runtime behaviour of the net evaluation. 

In this paper, we demonstrated how to encode interaction nets in OCaml manually. To allow for better comparisons with other tools, a compiler could be developed, taking as input a language similar to that of previous interaction net evaluators. OCaml offers an inbuilt system for language extension, therefore by developing an extension for interaction nets, interaction net algorithms could be embedded directly into a surrounding program, while automatically using parallel evaluation if run on a multicore platform. 

Finally, we want to remark, that while our focus here was on an encoding in OCaml, our method is general enough to be used for any language that offers GADTs and concurrency. 

\paragraph{Acknowledgements} 

This work is partially supported by the ERC CUSTOMER project. 

%% file: parts/app.tex
\section{Overview of OCaml}
\label{app:ocaml}

The basic units of any OCaml program are \emph{expressions}, such as \mintinline{ocaml}{10 + 25}. Each expression has a \emph{type}, such as int, float, bool, string, etc., and can be bound to a name via a \texttt{let} directive. The OCaml compiler uses type inference, so that for most ordinary cases no type annotations are needed. However, a programmer can always express the types of symbols explicitly if desired. The following definitions are equivalent: 

\begin{ocamlcode}
let x = 10 + 25 
let x : int = 10 + 25 
\end{ocamlcode}

Anonymous functions can be implemented with the \texttt{fun} keyword. As functions are first class values, they can also be bound to a name. Function types are indicated with the \texttt{->} symbol. The following definitions are all semantically equivalent: 

\begin{ocamlcode}
let square = fun x -> x * x 
let square : int -> int = fun x -> x * x 
let square x = x * x 
let square (x : int) : int = x * x 
\end{ocamlcode}

Functions are curried, so partial application is possible: 

\begin{ocamlcode}
let add x y = x + y 
let add2 = add 2 
let sum = add2 4 
\end{ocamlcode}

The core of OCaml's expressive power comes from the ability to define rich data types. The one most used in this paper is called \emph{variant data type} (a generalisation of enumeration and union types): 

\begin{ocamlcode}
type myType = 
    | A of int 
    | B of bool 
\end{ocamlcode}

The above expression introduces a type \texttt{myType}. Values of \texttt{myType} are \emph{either} an \texttt{A} carrying an integer as an attribute, \emph{or} a \texttt{B} carrying a boolean. Variants are constructed by applying their constructor name (e.g., \texttt{A} or \texttt{B}) to the appropriate number of argument expressions according to the type definitions. Variants can be deconstructed by \emph{pattern matching}: 

\begin{ocamlcode}
let x : myType = A 1
let output = match x with 
    | A n -> if n = 1 then true else false 
    | B b -> not b 
\end{ocamlcode}

We can see that the value carried by a constructor can be bound to a name on the left-hand side of a case, and then be referred to in the expression on the right-hand side. All cases of a pattern match must return the same type. Matches are also checked for exhaustiveness, so if we change the above example to

\begin{ocamlcode}
let x : myType = A 1
let output = match x with 
    | A n -> if n = 1 then true else false 
\end{ocamlcode}

the compiler will complain: 

\begin{textcode}
Warning 8 [partial-match]: this pattern-matching is not exhaustive.
Here is an example of a case that is not matched:
B _
\end{textcode}

Variants can be parameterized in terms of other types. For example, the OCaml standard library defines the list type in the following way: 

\begin{ocamlcode}
type 'a list = 
    | Nil 
    | Cons of 'a * 'a list 
\end{ocamlcode}

The \emph{type parameter} \texttt{'a} can be instantiated to any already defined type. Type synonyms can be introduced to make the code more readable: 

\begin{ocamlcode}
type myList = myType list 
let l : myList = Cons (A 1, Cons (B false, Nil)) 
\end{ocamlcode}

Type constructors are not functions, so their application cannot be partial.